%%%%%%%%%%%%%%%%%%%%%%%%%%%%%%%%%%%%%%%%%%%%%%%%%%%%%%%%%%%%%%%
\input harvmac
\input epsf
\overfullrule=0pt
\parskip=0pt plus 1pt
\sequentialequations

\def\ZZ{\hbox{\bf Z}}
\def\tr{\hbox{\rm tr}\,}

\newcount\figno
\figno=0
\def\fig#1#2#3{
\par\begingroup\parindent=0pt\leftskip=1cm\rightskip=1cm\parindent=0pt
\baselineskip=11pt
\global\advance\figno by 1
\midinsert
\epsfxsize=#3
\centerline{\epsfbox{#2}}
\vskip 12pt
\centerline{Fig.\ \the\figno: #1}\par
\endinsert\endgroup\par
}
\def\figlabel#1{\xdef#1{\the\figno}}
\def\np{Nucl.\ Phys.}
\def\pl{Phys.\ Lett.}
\def\prl{Phys.\ Rev.\ Lett.}
\def\cmp{Comm.\ Math.\ Phys.}

\def\pr{Phys.\ Rev.}
\def\jhep{JHEP} 

\Title{\vbox{\rightline{HRI-P/0312-001}
\rightline{TUW--03--39}
\rightline{hep-th/0312245}}}
{NS Fivebrane and Tachyon Condensation}

\medskip

\centerline{
Debashis Ghoshal$^a$,
Dileep P. Jatkar$^a$ and Maximilian Kreuzer$^b$}
\bigskip
\centerline{\it $^a$ Harish-Chandra Research Institute}
\centerline{\it Chhatnag Road, Jhusi, Allahabad 211019, India}
\smallskip
\centerline{\it $^b$ Institut f\"ur Theoretische Physik}
\centerline{\it Technische Universit\"at Wien}
\centerline{\it Wiedner Hauptstra\ss e 8-10, 1040 Vienna, Austria}
\smallskip
\centerline{\tt ghoshal, dileep@mri.ernet.in}
\centerline{\tt kreuzer@hep.itp.tuwien.ac.at}
\smallskip

\vglue .3cm
\vskip 2cm
\bigskip

\noindent We argue that a semi-infinite D6-brane ending on
an NS5-brane can be obtained from the condensation of the
tachyon on the unstable D9-brane of type IIA theory. The
construction uses a combination of the descriptions of
these branes as solitons of the worldvolume theory of the 
D9-brane. The NS5-brane, in particular, involves a gauge
bundle which is operator valued, and hence is better thought
of as a gerbe.

\Date{12/03}

% References
\def\STROM-TOWN{A.\ Strominger, {\it Open $p$-branes}, \pl\
{\bf B383} (1996) 44 [{\tt hep-th/9512059}];\hfill\break
P.\ Townsend, {\it Brane surgery}, \np\ (Proc.\ Suppl.)
{\bf 58} (1997) 163 [{\tt hep-th/9609217}].}
\def\HAWI{A.\ Hanany and E.\ Witten, {\it Type IIB superstrings,
BPS monopoles and three dimensional gauge dynamics}, \np\
{\bf B492} (1997) 152 [{\tt hep-th/9611230}].}
\def\GRHUTO{M.\ Green, C.\ Hull and P.\ Townsend, {\it D-brane
Wess-Zumino actions, T-duality and the cosmological constant},
\pl\ {\bf B382} (1996) 65 [{\tt hep-th/9604119}].}
\def\FIELDTH{D.\ Eardley, G.\ Horowitz, D.\ Kastor and J.\
Traschen, {\it Breaking cosmic strings without monopoles},
\prl\ {\bf 75} (1995) 3390;\hfill\break
A.\ Achucarro, R.\ Gregory and K. Kuijken, {\it Abelian
Higgs hair for black holes}, \pr\ {\bf D52} (1995) 5729 
[{\tt gr-qc/9505039}];\hfill\break
R.\ Gregory and M.\ Hindmarsh, {\it Smooth metrics for snapping 
strings}, \pr\ {\bf 52} (1995) 5598 [{\tt gr-qc/9506054}].}
\def\SENCON{A.\ Sen, {\it Non-BPS states and branes in 
string theory}, [{\tt hep-th/9904207}], and references
therein.}
\def\HORAVAK{P.\ Ho\v rava, {\it Type IIA D-branes, K-theory 
and matrix theory}, Adv.\ Theor.\ Math.\ Phys.\ {\bf 2} 
(2000) 1373 [{\tt hep-th/9812135}].}
\def\TORSION{E.\ Witten, {\it D-branes and K-theory}, \jhep\
{\bf 9812} (1998) 019 [{\tt hep-th/9810188}];\hfill\break
{\it Overview of K-theory applied to strings}, [{\tt 
hep-th/0007175}];\hfill\break
A.\ Kapustin, {\it D-branes in a topologically nontrivial
$B$-field}, Adv.\ Theor.\ Math.\ Phys.\ {\bf 4} (2000) 127
[{\tt hep-th/9909089}].}
\def\BOWMAT{P.\ Bouwknegt and V.\ Mathai, {\it D-branes,
B-fields and twisted K-theory}, \jhep\ {\bf 0003} (2000)
007, [{\tt hep-th/0002023}].}
\def\NCSOL{R.\ Gopakumar, S.\ Minwalla and A.\ Strominger,
{\it Noncommutative solitons}, \jhep\ {\bf 0005} (2000) 020
[{\tt hep-th/0003160}].}
\def\NCTACH{K.\ Dasgupta, S.\ Mukhi and G.\ Rajesh, {\it 
Noncommutative tachyons}, \jhep\ {\bf 0006} (2000) 022 
[{\tt hep-th/0005006}];
\hfill\break
J\ Harvey, P.\ Kraus, F.\ Larsen and E.\ Martinec, {\it
D-branes and strings as non-commutative solitons}, \jhep\
{\bf 0007} (2000) [{\tt hep-th/0005031}].} 
\def\HARMOOK{J.\ Harvey and G.\ Moore, {\it Non-commutative
tachyons and K-theory}, J.\ Math.\ Phys.\ {\bf 42} (2001) 
2765 [{\tt hep-th/0009030}].}
\def\EGKRS{S. Elitzur, A.Giveon, D. Kutasov, E. Rabinovici and
G. Sarkissian, {\it D-branes in the background of NS five-brane},
\jhep\ {\bf 0008} (2000) 046 [{\tt hep-th/0005052}].}
\def\CHANETAL{C.\ Chan, C.\ Chen and H. Yang, {\it Topological
$\hbox{\bf Z}_{N+1}$ charges on fuzzy sphere}, [{\tt
hep-th/0106269]}.}
\def\BRYLIN{J-L.\ Brylinski, {\it Loop spaces, characteristic classes
and geometric quantization}, Birkhauser (1993).}
\def\HTOPO{J.\ Harvey, {\it Topology of the gauge group in 
noncommutative gauge theory} in {\it Strings 2001}, Eds.\ A.\ 
Dabholkar {\it et al}, AMS (2002), [{\tt hep-th/0105242}].}
\def\NEAR{C.\ Callan, J.\ Harvey and A.\ Strominger,
{\it Worldsheet approach to heterotic instantons and
solitons}, \np\ {\bf B359} (1991) 611;\hfill\break
{\it Supersymmetric string solitons}, in Proc.\ of Trieste
Summer School (1991) [{\tt hep-th/9112030}].}
\def\KALK{J.\ Kalkkinen, {\it Gerbes and massive type II
configurations}, \jhep\ {\bf 9907} (1999) 002 
{[hep-th/9905018}].}
\def\HITCHIN{N.\ Hitchin, {\it Lectures on special lagrangian
submanifolds}, [{\tt math.DG/9907034}].}
\def\FRNEPO{P.\ Freund and R.\ Nepomechie, {\it Unified 
geometry of anti-symmetric tensor fields and gravity}, \np\
{\bf B199} (1982) 482.}
\def\GERBES{A.\ Carey, J.\ Mickelsson and M.\ Murray,
{\it Bundle gerbes applied to quantum field theory},
Rev.\ Math.\ Phys.\  {\bf 12} (2000) 65
[{\tt hep-th/9711133}];\hfill\break
C.\ Ekstrand,
{\it $k$-gerbes, line bundles and anomalies},
JHEP {\bf 0010} (2000) 038 [{\tt hep-th/0002063}];\hfill\break
Y.\ Zunger,
{\it p-gerbes and extended objects in string theory},
[{\tt hep-th/0002074}];\hfill\break
K.~Gawedzki and N.~Reis,
{\it WZW branes and gerbes},
Rev.\ Math.\ Phys.\  {\bf 14}, 1281 (2002)
[{\tt hep-th/0205233}];\hfill\break
J.~Mickelsson,
{\it Gerbes, (twisted) K-theory, and the supersymmetric WZW model},
[{\tt hep-th/0206139}].}
\def\RAERO{I.\ Raeburn and J.\ Rosenberg, {\it Crossed products
of continuous trace $C^*$-algebras by smooth actions},
Trans.\ Amer.\ Math.\ Soc.\ {\bf 305} (1988) 1 (Example 3.7).}
\def\ROSE{
J.\ Rosenberg, {\it Homological invariants of extension of 
$C^*$-algebras}, PSPM {\bf 38} (1982) 35.}
\def\BOTTU{R.\ Bott and L.\ Tu, {\it Differential forms in algebraic
topology}, Springer (1982).}
\def\ALCECH{O.\ Alvarez, {\it Topological quantization and
cohomology}, \cmp\ {\bf 100} (1985) 279.}
\def\KUI{N.\ Kuiper, {\it The homotopy type of unitary group of Hilbert
spaces}, Topology {\bf 3} (1965) 19.}
\def\NASH{C.\ Nash, {\it Differential topology and quantum field 
theory}, Academic Press (1991).}

\def\ALSCH{A.\ Alekseev and V.\ Schomerus, {\it D-branes in the WZW 
model}, \pr\ {\bf D60} (1999) 061901 [{\tt hep-th/9912193}].}
\def\STANC{S.\ Stanciu, {\it D-branes in group manifolds}, \jhep\
{\bf 0001} (2000) 025 [{\tt hep-th/09909163}].}
\def\MMS{J.\ Maldacena, G.\ Moore and N.\ Seiberg, {\it Geometrical 
interpretation of D-branes in gauged WZW models}, \jhep\ {\bf 0107}
(2001) 046 [{\tt hep-th/0105038}].}

{\nopagenumbers

\ftno=0
}
%%%%%%%%%%%%%%%%%%%%%%%%%%%%%%%%%%%%%%%%%%%%%%%%%%%%%%%%%%%%%%%

\newsec{Introduction}
In type IIA string theory a D6-brane can end on a Neveu-Schwarz
fivebrane in a supersymmetric
configuration\ref\StromTown{\STROM-TOWN}. The simplest way to 
see this is to start from a fundamental string ending on a 
D5-brane in type IIB theory. Indeed this defines a {\it Dirichlet} 
fivebrane. Now by S-duality followed by T-dualities along all 
the spatial directions of the worldvolume of the resulting 
NS5-brane we reach the desired configuration. 
Recall that this system (and its T-dual cousins) are 
essential ingredients in `brane engineering' of gauge theory
dynamics following Ref.\ref\HaWi{\HAWI}.

Naively
a semi-infinite brane in a flat space cannot exist by charge
conservation. There is a quantized charge of a D6-brane through 
a two-sphere enclosing it. However, in case of a semi-infinite
brane this $S^2$ can just be `slipped off' the end and collapsed,
leading to an apparent contradiction. This argument fails because
we are in a situation with a non-trivial Neveu-Schwarz $B$-field
provided by the fivebrane. The gauge-invariant field strength is
not simply the curvature of the RR one-form gauge 
field\ref\GrHuTo{\GRHUTO}. The NS $B$-field also couples to the 
Chan-Paton gauge field modifying the Bianchi
identity\ref\torsion{\TORSION} to $dF\sim H$. 

We would like to obtain the semi-infinite D6-brane ending on an
NS5-brane via tachyon condensation on the unstable D9-brane.
According to Sen\ref\SenCon{\SENCON}, all D-branes
in type IIA string theory arise as solitons of the worldvolume
theory on the D9-branes. In particular, the stable D6-brane is
an 't Hooft-Polyakov monopole of the gauge field-tachyon system 
on at least two D9-branes\ref\horavak{\HORAVAK}. The situation 
is more complicated in the presence of the non-trivial $B$-field 
due to the fivebrane
(see \ref\egkrs{\EGKRS} where many configurations involving 
D-branes and the NS5-brane were discussed). Any configuration must 
satisfy the modified Bianchi identity. In our case, the relation 
$dF\sim H$ must hold for both the final D6-NS5-brane configuration 
{\it after} the tachyon condensation, as well as {\it before} it, 
for the D9-NS5-brane system. The problem of tachyon condensation 
in presence of an $H$ field whose quantized charges are 
$\hbox{\bf Z}_n$ valued, was analyzed in Refs.\torsion. This was 
generalized to the usual integrally quantized case in 
Ref.\ref\BowMat{\BOWMAT}, which argues that in this situation one 
needs to consider the group of unitary operators in a Hilbert space 
as the gauge group on the D9-brane. Operator valued gauge fields 
appear in a natural way in the solitons of non-commutative 
gauge theory\ref\ncsol{\NCSOL}\ref\nctach{\NCTACH}. Indeed,
Harvey and Moore\ref\HarMooK{\HARMOOK} have suggested a configuration 
to describe an NS5-brane as a non-commutative soliton of  
open string theory (see also \ref\Chanetal{\CHANETAL}). We
present our arguments in this set-up.

It turns out that our construction is related to one version of
what is called a {\it gerbe}\ref\hitchin{\HITCHIN}, one in which 
it is described by operator valued gauge fields. (Let us note 
parenthetically that in their study of the antisymmetric tensor 
gauge fields, Freund and Nepomechie\ref\FrNepo{\FRNEPO} discovered 
gerbes in string theory. Some recent applications to string theory 
may be found in \ref\kalk{\KALK}\ref\gerbes{\GERBES}.) As 
a matter of fact, the NS5-D6-brane configuration has been 
obtained as a stable solution of massive type IIA 
supergravity\kalk, in the language of gerbes. However, 
a different description of gerbes in terms of local U(1) bundles, 
was used in \kalk, which did not discuss tachyon condensation either.

%%%%%%%%%%%%%%%%%%%%%%%%%%%%%%%%%%%%%%%%%%%%%%%%%%%%%%%%%%%%%%%

\newsec{Field theory analogue} 
It is instructive to look at a simpler field theory model, in
four spacetime dimension, which
share the essential features of the brane configuration we
wish to obtain after tachyon condensation.
This model consists of a semi-infinite Nielsen-Olesen vortex of
the abelian Higgs model ending on a Dirac monopole. We can think
of this monopole as a singular limit of the 't Hooft-Polyakov
monopole in the $SO(3)$ Georgi-Glashaw model. With the Higgs field 	%%the
pointed radially outwards in the field space, this is a
non-singular solution to the equations of motion with mass
proportional to $M_W/g_{YM}^2$. The Nielsen-Olesen vortex, on the
other hand, has constant finite energy per unit length. A
semi-infinite vortex ending on a monopole is then an infinite
energy configuration. To minimize its energy, the vortex will reduce
its length thereby pulling the monopole along all the way to
infinity. Hence the semi-infinite vortex string ending on a 
monopole is unstable.
 
There exists a remarkable way to stabilize this configuration
by putting the monopole inside an accelerating 
black hole\ref\fieldth{\FIELDTH}.
Here we begin with the abelian Higgs model coupled to gravity.
This model has cosmic string, i.e., Nielsen-Olesen vortex solution
as well as, say, a Schwarzschild black hole solution. Let us consider 
a configuration in which the vortex ends on a black hole. 
%%%%%%%%%%%%%%%%%%%%%%%%%%%%%%%%%%%%
In this case
one finds an axisymmetric metric with a conical singularity on the
accelerating Schwarzschild black hole, whose metric up to a conformal
factor is (see third reference in \ref\fieldth{\FIELDTH}),
\eqn\scmetric{
ds^2 = \left( 1-{2M\over r}-A^2r^2\right)dt^2 - \left( 1-{2M\over
r}-A^2r^2\right)^{-1}
dr^2 -r^2d\theta^2 - r^2(1-\alpha)^2\sin^2\theta d\phi^2,
}
where, $A$ is acceleration of the black hole.
The deficit angle of the conical singularity is proportional to
$\alpha$.
%%%%%%%%%%%%%%%%%%%%%%%%%%%%%%%%%%%%  Phys. Rev. 52(1995)5598 [gr-qc/9506054]
This is a reflection of the fact that the vortex is piercing the 
black hole horizon. The Schwarzschild black hole horizon has the
topology of a two sphere. Suppose in the frame of an asymptotic 
observer the vortex ends on the south pole of the horizon, then we 
can take a loop on the horizon around the south pole and measure the
magnetic flux of the vortex. For one vortex configuration, the {\it
angle valued} Higgs field $\Phi$ winds once around this loop. Now, 
by deforming this loop we can shrink it at the north pole. 
Since the vortex pierces the horizon only once, the shrinking 
of the loop at the north pole seems to lead to a contradiction. 
However, this result is misleading as the value of $\Phi$ is gauge 
dependent. The vortex is also accompanied by a topologically 
nontrivial gauge field configuration. A consistent 
solution corresponds to defining the Hopf 
fibration over $S^2$. This can be achieved by defining two
charts on $S^2$
\eqn\patches{
{\cal U}_N = \{\theta, \phi\, :\, \theta < \pi\},\quad\quad
{\cal U}_S = \{\theta, \phi\, :\, \theta > 0\}.
}
On the overlap, the fields are related by the transition function 
$g_{NS} = \exp{(- i\phi)}$ as
\eqn\trans{
\exp(i\Phi_N) = g_{NS}\,\exp(i\Phi_S), \qquad A_{N\mu} = A_{S\mu} + 
i g_{NS}^{-1}\partial_{\mu} g_{NS}.
}
Since the vortex is on the lower hemisphere, we can take
$\Phi_N = 0$ and $A_N =0$ on the northern hemisphere and connect 
it to the vortex configuration on the southern hemisphere via the
topologically nontrivial transition function. This 
configuration makes sense as long as the $S^2$ horizon does not 
shrink to zero size. This is ensured by the fact that the extended 
Schwarzschild geometry in the Kruskal coordinates is a wormhole 
with topology $S^2\times\hbox{\bf R}$, the minimum radius of the 
sphere being $2GM$. In the extended geometry, absence of the vortex 
in the northern hemisphere can be explained in the following manner. 
As the vortex approached the south pole of the horizon it goes down 
the throat of the wormhole and reappears through the horizon in 
the other asymptotic region.

Let us now get back to the winding number of $\Phi$. This
quantity is not gauge invariant in the presence of a nontrivial 
gauge field configuration. The field $\Phi$ is also 
not single valued everywhere on the sphere. We can, however, 
define a quantity 
\eqn\invar{
{\cal A} =  d \Phi - A,
}
which is both gauge invariant and single valued. In the northern 
hemisphere we have chosen $\Phi_N = 0$ and $A_N =0$, which implies 
even in the southern hemisphere the net winding charge
should vanish. Clearly ${\cal A}$ being single valued has zero 
winding charge. On the other hand, we saw that a vortex
solution near the south pole has $d\Phi$ winding number one. 
Hence we conclude that the integral of $A$ around the loop near the
south pole also has unit winding charge.

%%%%%%%%%%%%%%%%%%%%%%%%%%%%%%%%%%%%%%%%%%%%%%%%%%%%%%%%%%%%%%%

\newsec{Neveu-Schwarz fivebrane in open string theory}
The NS5-brane is a soliton of the closed string theory. Therefore 
one does not expect to see detailed features of it in open string
theory. Nevertheless, it turns out that some topological aspect
of the NS5-brane can be captured in terms of open strings. Harvey 
and Moore \HarMooK\ have a configuration with the right $H$-flux. In fact
they have argued that the NS5-brane may be thought of as a 
particular soliton in the non-commutative gauge theory of the
unstable D9-brane in type IIA theory. This is inspired by the
idea of a non-commutative tachyon\nctach. 

We will now review this construction. In \HarMooK, 
the spacetime topology is chosen to be 
$\hbox{\bf R}^{1,4}\times\hbox{\bf R}_{NC}^2\times S^2\times S^1$
and the $H$-flux is through the 3-cycle $S^2\times S^1$. 
This is based on an example in \ref\brylin{\BRYLIN}.
We will work, however, with 
$\hbox{\bf R}^{1,4}\times\hbox{\bf R}_{NC}^2\times S^3$, which
is the spacetime topology, at least in the near horizon limit, 
of the NS5-brane\ref\near{\NEAR}. 
There is a {\it constant} NS $B$-field along $\hbox{\bf R}_{NC}^2$. 
It should be emphasized that this is not the $B$-field that 
contributes to the $H$-flux, but has the effect of the 
making (the $\hbox{\bf R}_{NC}^2$ part of) spacetime non-commutative.
Therefore we may treat the tachyon, gauge and other fields as
opertator valued on $\hbox{\bf R}^{1,4}\times S^3$. Henceforth
we will concentrate only on the $S^3$ part.  

In Refs.\BowMat\ref\Htopo{\HTOPO}, it was argued that the gauge 
group in non-commutative gauge theory is 
$\hbox{\rm U}_{\hbox{\rm cpt}}({\cal H})$, a subgroup of unitary
operators in a Hilbert space ${\cal H}$ of the form 
$u=\hbox{\bf 1}+K$, where $K\in{\cal K}({\cal H})$ is a compact
operator. Conjugation by elements of the group U$({\cal H})$ 
are automorphisms of the corresponding Lie algebra 
${\cal K}({\cal H})$ of compact operators. However, since the 
U(1) centre of U(${\cal H}$) acts trivially, the automorphism of 
${\cal K}({\cal H})$, (and hence of 
$\hbox{\rm U}_{\hbox{\rm cpt}}({\cal H})$), is really 
$\hbox{\rm PU}({\cal H})=\hbox{\rm U}({\cal H})/\hbox{\rm U}(1)$.
The Lie algebra valued gauge field and the tachyon which 
transform in the adjoint representation are therefore valued in 
${\cal K}({\cal H})$. A non-trivial gauge bundle may be
constructed with a twist by an element of 
$\hbox{\rm Aut}({\cal K}({\cal H}))= \hbox{\rm PU}({\cal H})$.
The proposal of Ref.\HarMooK\ is that an appropriate non-trivial 
PU$({\cal H})$ bundle on $S^3$, with the tachyon field at the
maximum of the potential, represents a D9-brane and an
NS5-brane. Moreover, when the tachyon condenses to a minimum
of the potential, there is only an NS5-brane as the D9-brane
ought to have disappeared according to Sen's conjecture\SenCon. 

The key to the construction of this PU$({\cal H})$ bundle is
the fact that $\pi_2(\hbox{\rm PU}({\cal H}))=\hbox{\bf Z}$. It
is then possible to patch together trivial bundles on local
coordinate charts of $S^3$ along their overlaps. This is a one
higher dimensional generalization of the construction of a
monopole on $S^2$, which %%% one that 
used the fact that 
$\pi_1(\hbox{\rm U}(1))=\hbox{\bf Z}$. As we have mentioned 
before, the base space in \HarMooK\ is $S^2\times S^1$. The
Hopf fibration $S^3\rightarrow S^2$ and the covering space   of the fiber %%%
{\bf R}$\rightarrow S^1$ defines a natural 
$S^1\times\ZZ$ bundle on it. This is embedded in 
PU$({\cal H})$ by lifting the circle coordinate to an angle 
valued position operator $\hat\Omega$ and its (integrally
quantized) conjugate momentum $\hat L$ satisfying a 
Heisenberg algebra 
\eqn\heisenberg{
[\hat\Omega,\hat L] = i\,\hat{\hbox{\sl 1}}.}

Let us review some details of this construction following
Ref.~\brylin. One starts with a principle U(1) bundle $P$(U(1))
over a base manifold $X$ and the universal covering space
{\bf R} of $S^1$. This defines a principle 
$P(\hbox{\rm U(1)}\times\hbox{\bf Z})$ bundle over the
base space $X\times S^1$. In the following we will consider
the specific example of the Hopf bundle $X=S^2$ and 
$P$(U(1))$=S^3$. Given a (vector) space $V$ on which the 
group $G$ acts, it is possible to define a fibre bundle 
$E_V\rightarrow X\times S^1$ associated to the principle 
$G$ bundle by the quotient $(P(G)\times V)/G$. The fibre 
of this bundle\foot{Recall that in the associated bundle
the sections coming from a quotient action of $G$ on
$G\times V$ are identified as $(s,v)\,\sim\,(sg,g^{-1}v)$
providing a twist.} is isomorphic to $V$. 
%(For more details, see
%Appendix B.) 
The frame bundle
and the tangent bundle to a manifold is an example of such
a pair. Another natural association is one in which $V$
is the Lie algebra $\hat{\hbox{\tt g}}$ of $G$ or any of its 
representations. The $G$ action on $V$, in turn, induces an
adjoint action on the space of linear operators ${\cal L}(V)$
on $V$. One can, therefore, construct an associated `bundle'
whose fibre is ${\cal L}(V)$ and the transition functions
$g_{ij}$ act by adjoint action, (with the centre acting
trivially). In particular, our objective will be to 
construct a bundle whose fibre consists of operators in the 
Hilbert space ${\cal H}$ of square integrable functions 
$L^2(S^1)$ of a projective representation of the Heisenberg 
group $H$, which is a central extension of the group 
$S^1\times${\bf Z}. 

In order to specify the bundle, it will be sufficient to give
a local trivialization over open sets ${\cal U}_i$, in which
we specify the Hilbert space of functions and provide the
transition functions. Let $(x=(\psi,\theta),\phi)$ be %%% are 
points in $S^2\times S^1$ and
$$
p:S^3\times\hbox{\bf R}\rightarrow S^2\times S^1
$$
be %%%is 
the projection map of the smaller bundle with which the
construction proceeds. The fibre $p^{-1}(x)$ is a circle 
$S^1_x$ without any fixed base point. The universal cover
$\tilde{S^1_x}\sim\hbox{\bf R}_x$ is ambiguous up to the 
cyclic group generated by $T$ which shifts the coordinate
of $\hbox{\bf R}_x$ by ($2\pi$ times) an integer. Consider 
the Hilbert space of functions
\eqn\aitchx{
{\cal H}_{(x,\phi)} = \left\{f:\hbox{\bf R}_x\to\hbox{\bf R}_x
\;|\; f(T(\xi)) = e^{i\phi}\cdot f(\xi),\, \tilde\xi\in
\hbox{\bf R}_x\right\}.}
Of course, the space depends on the choice of coordinates on
the universal cover, but the ambiguity is upto an action of 
$T$, which acts as multiplication by a scalar leading to a
unique projective Hilbert space. Let $S^1$ be a circle which 
we can identify with the fibre $S^1_x$. With an abuse of
notation we will use the same angular coordinate $\xi$ on both 
these circles. Consider the square integrable functions $f(\xi)$ 
on the circle satisfying the following property under the 
isomorphism $\lambda_{(x,\phi)}:L^2(S^1)\sim
{\cal H}_{(x,\phi)}$
$$
\lambda_{(x,\phi)}\left(f(\xi)\right) = 
\exp\left({i\over 2\pi}\,\xi\phi\right)\,f(\xi).
$$
This specifies the local trivialization. It is easy to check 
that the above satisfies the property required of 
${\cal H}_{(x,\phi)}$ defined above. 

The Lie algebra $\hat{\hbox{\tt h}}$ of the group $H$ has 
generators $(\hat L,\hat{\hbox{\sl 1}})= \left(-i{d\over d\xi},
\hat{\hbox{\sl 1}}\right)$. Since the spectrum of the angular 
momentum $\hat L$ is discrete, there is no Lie algebra 
associated with the generator $\hat\Omega$ of the Heisenberg 
algebra \heisenberg. Rather $\exp(2\pi i\hat\Omega)\sim T$ 
is the generator of the automorphism discussed earlier. The
group $H$ acts by adjoint action on $\hat{\hbox{\tt h}}$.
While the action of the centre and the shift in $S^1$  
generated by $\hat L$ is trivial, the {\bf Z} acts 
nontrivially as follows:
\eqn\adjaction{
e^{2\pi i\ell\hat\Omega}\,\hat L\,
e^{-2\pi i\ell\hat\Omega} = \hat L - 2\pi\ell\, 
\hat{\hbox{\sl 1}}.}
The sections of the associated PU(${\cal H}$) bundle are
vectors $v(x,\phi)$ in $\hat{\hbox{\tt h}}$ satisfying
$$
v(x,\phi+2\pi\ell) = e^{2\pi i\ell\hat\Omega}\,v(x,\phi)\,
e^{-2\pi i\ell\hat\Omega}.
$$
Writing $v$ in terms of the basis elements as
$v=v_1(x,\phi)\hat{\hbox{\sl 1}}+v_2(x,\phi)\hat L$, we 
get: $v_1(x,\phi+2\pi\ell)=v_1(x,\phi)-2\pi\ell v_2(x,\phi)$
and $v_2(x,\phi+2\pi\ell)=v_2(x,\phi)$.

The final ingredient is a linear function from the 
PU(${\cal H}$) bundle to {\bf R}. In order to motivate this, let
us start with the exact sequence of vector spaces $V_C$ (generated
by $\hat{\hbox{\sl 1}}$), $\hat{\hbox{\tt h}}$ and $V_{\hat L}$
(generated by $\hat L$). The exact sequence of bundles
$$
E_{\hbox{\bf R}}\longrightarrow E_{\hat{\hbox{\tt h}}}
\longrightarrow E_{V_{\hat L}}
$$
follows from it, moreover, $E_{\hbox{\bf R}}\sim \hbox{\bf R}
\times(S^2\times S^1)$ is a trivial bundle. However, while in 
the former sequence $\hat{\hbox{\tt h}}$ cannot be written 
as a direct sum of $V_C$ and $V_{\hat L}$, it is possible to 
do so in the latter (although the Lie algebra will not be 
respected in the process). The linear function, which we will 
call `tr'
$$
\tr \,:\, E_{\hat{\hbox{\tt h}}} \rightarrow {\hbox{\bf R}},
$$
provides this decomposition. For a vector $v\in\hat{\hbox{\tt h}}$,
which can be written in terms of the basis elements as
$v=v_1(x,\phi)\hat{\hbox{\sl 1}}+v_2(x,\phi)\hat L$, we define
\eqn\deftrace{
\tr (v) = \left(v_1(x,\phi)-\phi\, v_2(x,\phi)\right). }
It is clear from \adjaction\ that the function tr is well defined 
on the PU(${\cal H}$) bundle. In particular, when $v_1=0$, $v_2=1$,
we have
$$
\tr (\hat L) = \phi.
$$
We will use this in a moment.

In order to specify a connection on this bundle, we start 
with a connection on the priciple U(1) bundle 
$p:S^3\rightarrow S^2$, which is the familiar monopole gauge 
field configuration $A^{(M)}$. The gauge field of this bundle is
a 1-form on $S^2$ valued in the Lie algebra generated by
$-i{d\over d\xi}$. The gauge connection of the PU(${\cal H}$)
bundle is a 1-form valued in $\hat{\hbox{\tt h}}$ and is 
taken to be $A^{(M)}$. Since $\hat{\hbox{\tt h}}$ is abelian,
the curvature of this connection is a 2-form 
$$
F^{(M)}=dA^{(M)}\,\hat L,
$$
where we have displayed the $\hat{\hbox{\tt h}}$ dependent part
explicitly\foot{This is same as writing $F=F^aT^a$ in YM theories.}. 
Acting with the linear function tr, we obtain
$\tr F^{(M)}=dA^{(M)}\,\phi$. This is called the `scalar 
curvature' in \brylin. It is a 2-form which is not closed,
rather $d\,\tr F^{(M)}\sim\hbox{\rm vol}(S^2\times S^1)$.

\bigskip

In case of the $S^3$ base, once again we use the Hopf fibration,
however, this time, following \ref\RaeRo{\RAERO}\ the U(1) 
action along the fibre is lifted
to a U(1) action in PU$({\cal H})$ with the help of a cocycle.
Let us consider $S^3$ as the unit sphere defined by 
$x_1^2+x_2^2+x_3^2+x_4^2=1$, which may be written as
$$
|z_0|^2 + |z_1|^2 = 1,
$$
in terms of complex coordinates of 
$\hbox{\bf C}^2[x_1+ix_2,x_3+ix_4]$. Let us cover $S^3$ by charts
\eqn\sphcharts{
\eqalign{
{\cal U}_0 &= \{(z_0,z_1)\in S^3 : |z_0|\ge |z_1|\}\cr
{\cal U}_1 &= \{(z_0,z_1)\in S^3 : |z_0|\le |z_1|\}.
}}
Each ${\cal U}_i$ is topologically a disc times $S^1$ and they 
overlap on a two-torus 
\eqn\overlap{
{\cal U}_0\cap{\cal U}_1 = T^2 = \{(z_0,z_1)\in S^3 : 
|z_0|=|z_1|=1/\sqrt{2}\}.}
The simplest way to see that $S^3$ has this topological structure
is to think $S^3=\hbox{\bf R}^3\cup\{\infty\}$. Now, if we remove
a solid torus from $\hbox{\bf R}^3$, what remains, together with 
the point at infinity, is also a solid torus. 
Introduce coordinates $(\psi,\theta,\phi)$ on $S^3$ such
that $z_0 = \cos{\psi\over2}e^{i(\phi+\theta)/2}$ and
$z_1 = \sin{\psi\over2}e^{i(\phi-\theta)/2}$. The bundle
structure of Hopf fibration is given by the local 
trivializations
\eqn\hopftrivial{
{\cal U}_0 \sim \left({z_1\over z_0},{z_0\over |z_0|}\right),\qquad
{\cal U}_1 \sim \left({z_0\over z_1},{z_1\over |z_1|}\right),
}
along with the transition function $z_0/z_1$ on the
overlap. The U(1) action 
\eqn\hopfuone{
(z_0,z_1)\;\rightarrow\; (e^{-i\omega}z_0,e^{-i\omega}z_1)
}
along the fibre is an isometry\foot{The metric in these
coordinates is 
$ds^2 = {1\over4}\left(d\psi^2 + \sin^2\psi\,d\theta^2 + 
(d\phi + \cos\psi\,d\theta)^2\right)$.} of $S^3$. 

The PU$({\cal H})$ bundle on $S^3$ is specified by a map
\eqn\pubundle{
g_{01}:{\cal U}_0\cap{\cal U}_1=T^2\;\rightarrow\; 
\hbox{\rm PU}({\cal H}).
}
Local trivializations
$$
f_0 : {\cal U}_0\rightarrow {\cal K(H)}\qquad\quad
f_1 : {\cal U}_1\rightarrow {\cal K(H)}
$$
are related by $f_0 = g_{01}(f_1) = g_{01} f_1 g_{01}^{-1}$ 
on the overlap. The 
topological properties of the bundle are %%%is 
characterized by 
the homotopy class of $g_{01}$, an element of maps from 
$T^2$ to PU$({\cal H})$. Now since 
$\pi_n(\hbox{\rm U}({\cal H}))=0$ for all 
$n$\ref\Kui{\KUI}, we have 
$\pi_{n-1}(\hbox{\rm U}(1))=\pi_n(\hbox{\rm PU}({\cal H}))$, 
hence the only non-vanishing homotopy group of PU$({\cal H})$ 
is $\pi_2$ and this is $\ZZ$. The homotopy classes of maps 
$g_{01}$ of interest is therefore isomorphic to 
$H^2(T^2,\ZZ)$ (see, for example, \ref\nash{\NASH}\ chapter 1). 
Now, using the relation\ref\bottu{\BOTTU} between the 
differential complexes on ${\cal U}_0,\, {\cal U}_1$, 
${\cal U}_0\cup{\cal U}_1=S^3$ and 
${\cal U}_0\cap{\cal U}_1=T^2$, we have  
$H^2(T^2,\ZZ)=H^3(S^3,\ZZ)=\ZZ$. 

Let us define the U(1) 
action \hopfuone\ on the PU$({\cal H})$ bundle as 
\eqn\puuone{
\eqalign{
\Omega_\omega\, :\, (f_0,f_1) &\rightarrow 
(f^\omega_0,f^\omega_1)\cr
\hbox{\rm where, }\qquad f^\omega_0(z_0,z_1) &=  
f_0(e^{-i\omega}z_0,e^{-i\omega}z_1)\cr
f^\omega_1(z_0,z_1) &=  h(\omega,z_0,z_1)\,
\left(f_1(e^{-i\omega}z_0,e^{-i\omega}z_1)\right), 
}}
for a function 
$h(\omega,z_0,z_1): S^1\times{\cal U}_1\rightarrow
\hbox{\rm PU}({\cal H})$ which satisfy
\eqn\cocyle{
\eqalign{
h(\omega,z_0,z_1) &= g_{01}^{-1}(z_0,z_1)\,
g_{01}(e^{-i\omega}z_0,e^{-i\omega}z_1)\quad\hbox{\rm for }
(z_0,z_1)\in{\cal U}_0\cap{\cal U}_1,\cr
h(\omega_1+\omega_2,z_0,z_1) &= h(\omega_1,z_0,z_1)\, 
h(\omega_2,e^{-i\omega_1}z_0,e^{-i\omega_1}z_1)
\quad\hbox{\rm for } (z_0,z_1)\in {\cal U}_1.
}}
The first of the conditions \cocyle\ ensures that the patching
condition given by \pubundle\ is respected. In other words, 
$\Omega_\omega$ is an automorphism of the triple 
$(f_0,f_1;g_{01})$ used to define the PU$({\cal H})$ bundle. The 
second condition is a group homomorphism that identifies an $S^1$ 
in PU$({\cal H})$. The existence of $\Omega_\omega$ with the 
required properties is proven in Ref.\RaeRo\foot{The 
PU$({\cal H})$ bundle on $S^3$ has also been 
defined through local trivializations $S^3=D_+^3\cup D_-^3$, 
$D_+^3\cap D_-^3=S^2$ at 
the equator\ref\Rose{\ROSE}. The topological
properties of the bundle are characterized by maps from $S^2$ to
PU$({\cal H})$. This, however, does not seem suitable for our 
purpose as there is no natural U(1) action.}.

We will use the above to propose a construction along the lines
of \HarMooK, with $\Omega_\omega$ playing the role of 
$\hat\Omega$ in \heisenberg. We cannot identify $\hat L$ explicitly
in PU(${\cal H}$), but proceed with the assumption that there is
one such $\hat L$ such that \heisenberg\ is true. This assumption
is not untenable since the size of the fibre in Hopf fibration is
fixed, namely $4\pi$, therefore the spectrum of $\hat L$ is
discrete. Motivated by the construction in $S^2\times S^1$ and
the Hopf bundle description of $S^3$ outlined above, we
propose the following expression for the gauge fields using 
the two charts \hopftrivial\ of $S^2$.
\eqn\puconnection{
\eqalign{
A_0 &= +{i\over2}(1-\cos\psi) d\theta\cdot\hat L\cr
A_1 &= -{i\over2}(1+\cos\psi) d\theta\cdot\hat L,
}}
where we have displayed the algebra generator explicitly.
Recall the charts overlap for $\psi={\pi\over2}$, where the
transition functions $(\psi,\theta)\rightarrow (\pi-\psi,-\theta)$ 
of the Hopf bundle ensures that $A_0$ and $A_1$ differ by a gauge 
transformation. 

The `scalar curvature' of the gauge field \puconnection\ is 
obtained by taking the tr:
$$
\tr F = {i\over2}\sin\psi\,d\psi\wedge d\theta\,(\phi+\theta),
$$
where, in analogy with \deftrace, we have used $\tr\hat L = 
\phi+\theta$, the value of the coordinate of the Hopf fibre 
of the base space. Notice that this %%%
is a well defined 2-form. Finally,
\eqn\dFeqH{
d\,\tr F = {i\over2}\sin\psi\, d\psi\wedge d\theta\wedge d\phi,}
is the volume form on $S^3$ yielding a unit 3-form flux
through it. It is assumed here that the tachyon
is trivial, it is zero corresponding to the maximum of the potential. 
In other words, this gauge field configuration describes the unstable 
D9-brane in presence of the NS5-brane.

The NS5-brane so constructed has its worldvolume along 
$\hbox{\bf R}^{1,4}$ as well as along one of the non-commutative
dimensions in $\hbox{\bf R}_{NC}^2$. We refer to \HarMooK\ for 
some subtleties with this description. 

Finally, although we have been talking about a PU$({\cal H})$
bundle, the above construction is not a bundle in the usual
sense. In Appendix A, we show how it satisfies the 
conditions required of a gerbe. 
 
%%%%%%%%%%%%%%%%%%%%%%%%%%%%%%%%%%%%%%%%%%%%%%%%%%%%%%%%%%%%%%%

\newsec{Semi-infinite D6-brane and NS fivebrane}
In the previous section we have constructed %%%
a configuration of the operator
valued gauge fields in the noncommutative worldvolume theory of
the unstable D9-brane of type IIA theory. This configuration
carries a unit $H$-flux through $S^3$, and satisfies the
modified Bianchi $d\,\tr F = H$. It is argued\HarMooK\ that
for the tachyon at the maximum of the potential, this 
configuration describes the D9-NS5-brane system, while at
a minimum there is only an NS5-brane. One expects that, when
the tachyon is non-trivial, we should have a
configuration of the NS5-brane together with a D-brane of 
appropriate codimension. 

Recall, that in the absence of any $H$-flux, (no NS5-brane), 
all the stable D$p$-branes may be obtained as odd codimension
soliton solutions of the tachyon and gauge field theory. In 
particular the D6-brane is the 't Hooft-Polyakov monopole 
of the U(2) theory on two D9-branes\horavak. Let 
$(x_1,x_2,x_3)$ be the space transverse to the would be D6-brane. 
Identifying SU(2)$\subset$U(2) with the (covering space) of
the SO(3) group of rotations, the configuration is
\eqn\tHP{
\eqalign{
T &\sim x_i\sigma^i\cr
A_i &\sim \epsilon_{ijk}x^j\sigma^k,
}}
where $\sigma^i$ are the Pauli matrices. There should also be
some convergence factors on the right hand side. One important
feature of this construction is that it is local, {\it i.e.}~it
relies only on coordinates in a small neighbourhood of the 
origin where the D6-brane is located. 

The configuration we would like to obtain is that of a 
semi-infinite D6-brane that ends on the NS5-brane. The fivebrane
shares all its worldvolume dimensions with the D6-brane, whose
additional dimension has a boundary on which the NS5-brane 
lies. Let us put the NS-brane at the origin of its transverse 
$\hbox{\bf R}^4$ directions. This space is foliated by $S^3$
of varying radii, with the size finally saturating to give 
the `throat' geometry\near. The D6-brane appears to be a string
which pierces the $S^3$'s at, say, the south pole. Although, to 
an observer far away from the origin, it would seem that the 
D6-brane ends on the NS5-brane at the origin, it would be more 
correct to say that it goes down the `throat'. Actually for our 
case, where there are two non-commutative dimensions and the 
D6-brane worldvolume extends along both; this picture is an
extrapolation from the commutative limit. In particular, the
radial direction in $\hbox{\bf R}^4$, {\it i.e.}, the direction
transverse to $S^3$ is one of the non-commutative directions.

We would like to argue that the situation is different 
{\it after} tachyon condensation. The operator $\hat\Omega$ 
is a shift along the Hopf fibre of $S^3$. The process of 
tachyon condensation selects a special point, which we may 
choose to be the south pole. The fields are localized around 
this point, in particular, also along the Hopf fibre through 
it. This in turn, determines the value of the operator 
$\hat\Omega$ to be, say, zero, to an accuracy 
$\Delta\Omega\sim\varepsilon$. As a result, there is a large 
uncertainty in the value of the conjugate variable $\hat L$:
$\Delta L\sim 1/\varepsilon$. This, in effect makes the spectrum
of $\hat L$ continuous for sufficiently small $\varepsilon$.
Therefore, after the tachyon has condensed, it should be 
possible to shift $\hat L$ by an arbitrary amount. This is
in contrast to the previous section in which the fields 
are not localized in $S^3$.

We can follow the field theory example in Sec.~2 and use 
continuous gauge transformation of the form 
$\exp(ix\hat\Omega)$ (for any real $x$) available now to 
propose the following gauge field configurations: 
\eqn\newgaugeconfig{
\eqalign{
A_0 &= +{i\over2}(1-\cos\psi) d\theta\cdot\hat L\cr
A_1 &= -{i\over2}(1+\cos\psi) d\theta\cdot
\left(\hat L + (\phi-\theta) \hat{\hbox{\sl 1}}\right). 
}}
At the overlap, an operator valued U(1) gauge transformation 
\eqn\bgaugetr{
A_1 = e^{-i(\phi-\theta)\hat\Omega}\,A_0\, 
e^{i(\phi-\theta)\hat\Omega},}
relates the gauge configurations from the chart ${\cal U}_0$ 
to ${\cal U}_1$.

An operator valued U(1) gauge transformation is in fact 
equivalent to the gauge transformation of the $B$ field 
(see Appendix A and \gerbes). 
The field configuration \newgaugeconfig\
has the property that in the chart ${\cal U}_0$,  
$d\,\tr F_0\sim\hbox{\rm vol}(S^3)$ as before, but in 
${\cal U}_1$, $\tr F_1$ and hence $d\,\tr F_1$ vanish. Hence 
there is an NS $H$-flux through ${\cal U}_0$ but none 
through ${\cal U}_1$.

Continuing to follow the field theory example, we now need 
to show that the operator valued U(1) gauge field
configuration \newgaugeconfig\ arises from  a configuration of
the tachyon and gauge fields (in the $S^3$ part of the
worldvolume of the non-BPS D9-brane). This ought to localize
the energy around the south pole of $S^3$, which we assume
is at the origin of ${\cal U}_0$. Unfortunately, we are not able 
to write this explicitly. However, the operator corresponding to 
the tachyon field is expected to be of the form
\eqn\tachguess{
T\sim\int d\varphi\;|\varphi\rangle\langle\varphi|\,
e^{-\varphi^2/\varepsilon^2},}
which is a projection operator. Interestingly, in \horavak, 
it is shown that the configuration \tHP\ can also be thought 
of as a two step process of a vortex and a kink. This seems
more natural in the present situation as the local symmetry
group around the south pole is SO(2)$\times${\bf R}, since
${\cal U}_0$ has the topology of a cylinder $D_0^2\times S^1$,
which naturally accomodates this break up. 

%%%%%%%%%%%%%%%%%%%%%%%%%%%%%%%%%%%%%%%%%%%%%%%%%%%%%%%%%%%%%%%

\newsec{Bianchi identities and charge conservation}
In Ref.\kalk\ the NS5-D6-brane configuration described in the
previous section was shown, following a construction in \hitchin,
to be a solution of type IIA supergravity using a description
of gerbes as local line bundles. There the various 
charge conservation conditions are discussed in details. 
Briefly, consider an $S^2$ surrounding the D6-brane. There is
a flux of the RR one-form gauge field $C^{(1)}_{RR}$ through it.
In the absence of any $H$-flux, this measures the quantized
RR charge of the D6-brane. This arises from the Chern-Simons
couplings, ($dT\wedge F_{CP}\wedge C^{(7)}$, 
$dT\wedge dT\wedge dT\wedge C^{(7)}$, etc), on the D9-brane. 
When an NS $B$-field is
present, the correct gauge invariant field strength for this 
field is\GrHuTo\ 
\eqn\RRF{
G^{(2)}_{RR} = d C^{(1)}_{RR} + m B,}
where, $B$ is the NS 2-form and $m$ is the mass parameter. 
Similarly, 
\eqn\CPF{
{\cal F} = d A_{CP} - B,}
is a gauge invariant combination of the field strength involving 
the Chan-Paton gauge field. 

In our description, the flux is through a 2-cycle $T^2$ at the 
overlap of ${\cal U}_0$ and ${\cal U}_1$. This is assumed
to enclose D6-brane at the south pole of $S^3$. Since the 
D6-brane is semi-infinite, there is no Chan-Paton gauge field 
flux through the chart ${\cal U}_1$. We can also choose to set 
$B=0$ here. This means ${\cal F}=0$ in ${\cal U}_1$, and
in particular, there is no ${\cal F}$-flux through it.
Notice that ${\cal F}$ is gauge invariant and therefore
${\cal F}$-flux must vanish everywhere. The overlap $T^2$ is 
the boundary of a 3-space ${\cal U}_0$, which is that part of 
the $S^3$ through which there is a nontrivial $H$-flux. The 
anomalous Bianchi identity\GrHuTo\ from \RRF\
ensures that there is no net six-brane charge. By drawing analogy
with the field theory example, it now follows that the net monopole 
charge through a $T^2$ (at the overlap of ${\cal U}_0$ and 
${\cal U}_1$ or any deformation of it in ${\cal U}_0$), enclosing
the south pole of $S^3$ should also vanish. Since ${\cal F}$-flux 
through $T^2$ vanishes and
the flux of $dA_{CP}$ does not, we conclude that the monopole 
charge evaluated by integrating $dA_{CP}$ over a $T^2$ enclosing
the south pole is equal to the boundary value of the $H$-flux 
through the 3-space $D_0^2\times S^1$ enclosed by $T^2$, i.e.,
\eqn\chrgnil{
\oint_{T^2} dA_{CP} = \displaystyle\int_{D_0^2\times S^1} H = 
\oint_{T^2} B.
}
This in effect implies a modified Bianchi identity
\eqn\mbi{
dF_{CP} = H
}
for the NS5-D6-brane configuration.

%%%%%%%%%%%%%%%%%%%%%%%%%%%%%%%%%%%%%%%%%%%%%%%%%%%%%%%%%%%%%%%

\newsec{Discussion}
We have argued how to realize a configuration in which a
semi-infinite D6-brane ends on an NS5-brane via condensation
of the tachyon field on the worldvolume of unstable D9-branes.
Both the five- as well as the six-brane are solitonic 
configurations in the non-commutative field theory on the
D9-brane. Let us emphasize that although the six-brane by itself
is a solution of this field theory, the NS5-brane is only a
configuration. In the framework of open string theory, it is
as yet unclear in what sense a solitonic object of closed string 
theory can be realized as a solution. Some topological aspect of 
the NS5-brane can, however, be reproduced. Our intersecting
brane configuration, in which we have combined features of
\horavak\ and \HarMooK, is also not a solution of the open
string equations of motion. 

The geometrical description we have used is strictly valid for 
large values of NS5-brane charge, or far away from the core of
the fivebrane. Moreover, two of the longitudinal directions 
of our D6-brane carry a constant $B$ field. This is the $B$ field 
introduced to have a non-commutative worldvolume theory. The 
Chern-Simons coupling $B\wedge C^{(5)}_{RR}$ therefore results 
in an induced D4-brane charge in the configuration. 

Let us end by some speculative remarks on D3-branes in SU(2) 
WZW model. In this case, it is well-known that the symmetries
allow only D2-branes and D0-branes along conjugacy classes
of the group manifold\ref\AlSch{\ALSCH}. These are $S^2$'s at 
some fixed `latitudes' of $S^3$. On the other hand, 
Ref.\ref\stanc{\STANC}\ argued in favour of D3-branes which
wrap almost all of $S^3$ except for a set of points. More
recently, based on consistency with T-duality,
Ref.\ref\mms{\MMS}\ showed that there should be D3-branes 
which are `fat' D-strings. These have the topology of a 
cylinder reminiscent of the coordinate charts we have used 
in our construction of the NS5-D9-brane configuration. While
a single fat string cannot cover the entire group manifold
without having a singularity, it seems possible for a 
configuration of two fat D-strings `linked' together to 
do so. A nontrivial linking should capture the fact that
this configuration is a gerbe. This may be possible with 
operator valued gauge fields on the fat strings.

\noindent{\bf Acknowledgement}:
It is a pleasure to acknowledge fruitful discussions with
Rajesh Gopakumar, Gautam Mandal, Dipendra Prasad,
Volker Schomerus, Ashoke Sen, David Tong, Spenta Wadia and 
especially Indranil Biswas, who patiently explained gerbes 
to us. The work of M.K. was supported in part by the FWF 
under grant Nr. P15553.

%%%%%%%%%%%%%%%%%%%%%%%%%%%%%%%%%%%%%%%%%%%%%%%%%%%%%%%%%%%%%%%

\appendix{A}{Gerbes}
Gerbes are generalization of U(1) bundles (more generally line
bundles) on a manifold. This appendix contains a quick description 
of gerbes. Further details and references can be found in the 
expository article by Hitchin\hitchin. Ref.\gerbes\ is an 
incomplete list of their applications in string theory. 

Consider a manifold ${\cal M}$ and a set of open charts 
$\{{\cal U}_i\}$ that covers it: ${\cal M}=\cup_i{\cal U}_i$. 
We will assume, for simplicity, that each ${\cal U}_i$ is
contractible. A {\it 1-gerbe} ${\cal G}$ on ${\cal M}$ is
defined by a set of U(1) bundles ${\cal L}_{ij}$ on each 
(ordered) overlap ${\cal U}_i\cap{\cal U}_j$, satisfying the 
following conditions
\item{(i)} ${\cal L}_{ji}={\cal L}_{ij}^*$, (where ${\cal L}^*$ 
is the bundle dual to ${\cal L}$),
\item{(ii)} on triple overlaps 
${\cal U}_i\cap{\cal U}_j\cap{\cal U}_k$, the tensor product
bundle ${\cal L}_{ij}\otimes{\cal L}_{jk}\otimes{\cal L}_{ki}$
has a nowhere vanishing section $s_{ijk}$,
\item{(iii)} on quadruple overlaps 
${\cal U}_i\cap{\cal U}_j\cap{\cal U}_k\cap{\cal U}_l$,
the section $s_{ijk}\otimes s^*_{ijl}\otimes s_{ikl}\otimes
s^*_{jkl}=1$. 

Notice that in the last condition, the tensor product of the 
sections is that of a trivial bundle 
${\cal M}\times\hbox{\rm U(1)}$, as follows from the other two
conditions. Let us also note that if we take 
${\cal L}_{ij}={\cal L}_i\otimes{\cal L}^*_j$, where 
${\cal L}_i$ are U(1) bundles on ${\cal U}_i$, then all the
conditions are trivially satisfied. Therefore this is called
a trivial gerbe.

The above may be generalized to $k$-gerbes by defining line
bundles on $(k+1)$-fold overlaps with appropriate conditions.
An ordinary line bundle is a 0-gerbe from this point of view. 
It should be noted that, (except for $k=0$), the `total space'
of a gerbe is not a manifold, as the definition involves 
conditions on more than two overlaps. 

A {\it connection} on a 1-gerbe is specified by connections
$A_{ij}$ for each ${\cal L}_{ij}$ and a two-form $B_i$ on 
each chart ${\cal U}_i$, such that
\item{(i)} $A_{ij}=-A_{ji}$,
\item{(ii)} $s_{ijk}$ is flat with respect to the induced 
connection,
\item{(iii)} on ${\cal U}_i\cap{\cal U}_j$, we have
$B_i - B_j = dA_{ij}$.

The `curvature' $H=dB$ of this connection is independent of the
chart and hence makes sense globally. The cohomology of $H$ is
characterized by $H^3({\cal M},\hbox{\bf Z})$. The quantization
is analogous to the case of usual U(1) gauge 
fields\ref\alcech{\ALCECH}.

As an example, let us describe the `NS5-brane'. Consider spacetime
of the form $\hbox{\bf R}^{1,5}\times\hbox{\bf R}\times S^3$, the
geometry of the NS5-brane. The only relevant part of it is $S^3$,
on which we will construct a gerbe such that it carries an 
$H$-flux. This example is due to Hitchin\hitchin\ (and has been
used in \kalk). First, we
cover $S^3$ with two open 3-discs $D^3_\pm$, which overlap 
around a region around the equatorial $S^2$. The overlap
$D^3_+\cap D^3_-$ has the topology of a `cylinder' 
$S^2\times\hbox{\bf R}$. For the U(1) bundle on the overlap,
we take the monopole bundle on $S^2$ (more precisely, 
the pull-back of this bundle). Let $A_{+-}$ be the gauge
field and $F=dA_{+-}$ be its curvature. In order to give their
concrete forms, let us introduce coordinates 
$(\alpha,\beta,\gamma)$ on $S^3$, such that the metric is
$$
ds^2 = d\alpha^2 + \sin^2\alpha\left(d\beta^2 + \sin^2\beta
d\gamma^2\right).
$$ 
The overlap region is ${\pi\over2}-\epsilon < \alpha <
{\pi\over2}+\epsilon$, and $F_{+-}\sim\sin\beta\,d\beta\wedge
d\gamma$. We need to specify the gerbe connections $B_\pm$. To
this end, consider a partition of unity $\varphi_\pm$. Recall 
that these are functions with supports respectively in $D^3_\pm$, 
such that $0\le \varphi_\pm\le 1$ and $\varphi_+ + \varphi_- =1$ 
at each point. We write,
$$
B_\pm = \pm\varphi_\pm F_{+-},
$$
which satisfy the condition $B_+ - B_- = F_{+-}$. It is easy to
check that the curvature $H=dB$ is independent of the chart. In
fact it equals $F_{+-}\wedge d\varphi_+$, which is supported
on the overlap. Therefore, using the quantization of the 
monopole field $F_{+-}$, we see that the $H$-flux through $S^3$ 
is integrally quantized. It is curious that the gerbe 
defining an NS5-brane is roughly like a monopole ($F$ part) 
times a kink ($\varphi$ part), quite similar to, say, the 
soliton description of D6-brane in \horavak. Ref.\kalk\
describes the NS5- and semi-infinite D6-brane configuration 
in this language. 

Finally, let us show how the ${\cal K}({\cal H})$ valued gauge
fields patched together by 
$\hbox{\rm Aut}({\cal K}({\cal H}))= \hbox{\rm PU}({\cal H})$
satisfy the axioms of a gerbe. In the construction Sec.3, we have 
only two coordinate charts, so there is not much to check. 
Consider, instead a general set up where, we have local
trivializations given by maps
$$
f_i : {\cal U}_i\rightarrow {\cal K(H)},
$$
which satisfy $f_i = g_{ij}(f_j) = g_{ij} f_j g^{-1}_{ij}$,
for $g_{ij}\in \hbox{\rm PU}({\cal H})$, on twofold overlaps. 
Hence, on triple overlaps $h_{ijk}=g_{ij}g_{jk}g_{ki}$ must 
be an element of U(1), (since action of the U(1) centre of 
U$({\cal H})$ is trivial). These $h_{ijk}$'s may be taken
as the sections $s_{ijk}$'s of (trivial) U(1) bundles on 
threefold overlaps. It is also easy to check that the 
conditions on fourfold overlaps is satisfied. 

%\appendix{B}{Associated bundles}
%Let us briefly recall the construction of an associated bundle. 
%We start with a principle $G$ bundle $P(G)$ over a base space
%$X$. Suppose $x\in X$ is a point that is common to two coordinate
%charts ${\cal U}_i$ and ${\cal U}_j$ and let $s_i(x)$ and $s_j(x)$ 
%be sections such that
%%%
%$$
%s_j(x)=s_i(x)g_{ij}(x),\qquad g_{ij}\in G.
%$$
%%%
%Consider a vector space $V$ on which $G$ acts as follows:
%%%
%$$
%\eqalign{
%G\; :\; V &\rightarrow V\cr
%v &\mapsto g v.}
%$$
%%%
%We start with the product of the $G$ bundle $P(G)$ with the
%trivial bundle $V\times X$. The following identification of
%the sections coming from a quotient action of $G$ on 
%$G\times V$ 
%%%
%$$
%(s,v)\,\sim\,(sg,g^{-1}v)
%$$
%provides a twist yielding a nontrivial bundle over $X$ whose 
%fibre is isomorphic to $V$. The sections $(s_i(x),v(x))$ and
%$(s_j(x),v(x))$ of the product can be related by the quotient
%as
%%%
%$$
%(s_j(x),v(x))=(s_i(x)g_{ij}(x),v(x))\sim
%(s_i(x),g_{ij}(x)v(x)),
%$$
%%%
%from which the twist of $V$ bundle is apparent. This bundle is
%called an associated bundle. The $G$ action on $V$ induces an
%adjoint action on the space of linear operators ${\cal L}(V)$ 
%on $V$. One can, therefore, construct an associated `bundle' 
%whose fibre is ${\cal L}(V)$ and the transition functions
%$g_{ij}$ act by adjoint action. 

\vfill\eject

\listrefs 

\bye